\documentclass[runningheads]{llncs}
\usepackage[T1]{fontenc}
\usepackage{graphicx}
\usepackage{subcaption}
\usepackage{amsmath}
\usepackage{booktabs}
\usepackage[table,xcdraw]{xcolor}
\usepackage[font=small]{caption}
\usepackage{algorithm}
\usepackage[noend]{algpseudocode}
\usepackage{amssymb} 
\usepackage{tabularx}
\usepackage{xspace}
\usepackage{array}
\usepackage{booktabs}
\usepackage{adjustbox,makecell}
\usepackage[inline]{enumitem}
\usepackage[pagebackref=false,breaklinks,colorlinks]{hyperref}
\usepackage{pifont} 
\usepackage{siunitx}

\usepackage[capitalize,nameinlink]{cleveref}
\crefformat{figure}{#2\figurename~(#1)#3}

\def\xray{\texttt{X-Ray}\xspace}
\def\dl{\texttt{DL}\xspace}
\def\hffed{\texttt{HF-Fed}\xspace}
\def\non{\texttt{NoN}\xspace}
\def\fl{\texttt{FL}\xspace}
\def\cl{\texttt{CL}\xspace}
\def\noniid{\texttt{Non-IID}\xspace}

\begin{document}

\title{HF-Fed: Hierarchical based customized Federated Learning Framework for X-Ray Imaging}

\titlerunning{\hffed}

\author{Tajamul Ashraf\inst{1}\textsuperscript{\textdagger}\orcidID{0000-0002-7372-3782} \and
Tisha Madame\inst{1}\orcidID{0009-0009-7190-6591}}

\institute{\textsuperscript{1}Indian Institute of Technology Delhi, New Delhi 110016, INDIA\\
\email{tajamul@sit.iitd.ac.in}}

\authorrunning{Tajamul et al.}

\renewcommand{\thefootnote}{\textdagger}
\footnotetext{Corresponding author}

\maketitle             

\begin{abstract}

In clinical applications, \xray technology plays a crucial role in noninvasive examinations like mammography, providing essential anatomical information about patients. However, the inherent radiation risk associated with \xray procedures raises significant concerns. \xray reconstruction is crucial in medical imaging for creating detailed visual representations of internal structures, and facilitating diagnosis and treatment without invasive procedures. Recent advancements in deep learning (\dl) have shown promise in \xray reconstruction. Nevertheless, conventional \dl methods often necessitate the centralized aggregation of substantial large datasets for training, following specific scanning protocols. This requirement results in notable domain shifts and privacy issues. To address these challenges, we introduce the Hierarchical Framework-based Federated Learning method (\hffed) for customized \xray Imaging. \hffed addresses the challenges in \xray imaging optimization by decomposing the problem into two components: local data adaptation and holistic \xray Imaging. It employs a hospital-specific hierarchical framework and a shared common imaging network called Network of Networks (\non) for these tasks. The emphasis of the \non is on acquiring stable features from a variety of data distributions. A hierarchical hypernetwork extracts domain-specific hyperparameters, conditioning the \non for customized \xray reconstruction. Experimental results demonstrate HF-Fed's competitive performance, offering a promising solution for enhancing \xray imaging without the need for data sharing. This study significantly contributes to the evolving body of literature on the potential advantages of federated learning in the healthcare sector. It offers valuable insights for policymakers and healthcare providers holistically.  The source code and pre-trained \hffed model is available at this \href{https://tisharepo.github.io/Webpage/}{link}
%
\end{abstract}

\section{Introduction}

\xray imaging is vital for clinical diagnosis, providing noninvasive insights into patient anatomy, especially in breast mammography~\cite{10.1093/jnci/djy222}, \cite{rodriguez2019detection}, \cite{geras2019artificial}. However, potential radiation risks related to genetic and cancer diseases have raised concerns~\cite{hasegawa1990physics}. Common strategies to mitigate these risks, like adjusting the \xray tube's current/voltage and reducing scanning views, often degrade imaging quality~\cite{slovis2002alara}, negatively impacting image analysis and diagnoses. To address these challenges, researchers have explored deep learning (\dl) for low-dose \xray (\texttt{LD-Xray}) reconstruction~\cite{ikuta2022deep}, with promising results. However, current methods depend on centralized training (\cl) with large datasets from various hospitals, overlooking privacy concerns and domain shifts that cause low accuracy. Anonymizing data has proven insufficient in ensuring robust patient privacy~\cite{narayanan2008robust,schwarz2019identification}.

In light of privacy, legal, and ethical concerns, there is an increasing demand for a collaborative, privacy-preserving approach in multi-hospital training. Federated Learning (\fl) is a paradigm designed to enhance data security and privacy by decentralizing the training process, keeping information on individual devices, and
sharing only the updated model parameters~\cite{li2020federated}, \cite{hanzely2020federated}. \fl stands out as a decentralized solution explicitly crafted to safeguard data and its confidentiality~\cite{kaissis2020secure}. Unlike \cl methods, which involve the transfer of private patient data, \fl methods exclusively exchange gradients, thereby minimizing privacy risks associated with data content. \fl offers a privacy-conscious alternative for training models on \xray data from multiple sources, addressing the limitations of centralized approaches and fostering a more secure and ethically sound framework for medical imaging.
A significant challenge in \fl arises from the non-independence and non-identically distributed (\noniid) nature of data, a particularly pronounced issue in \xray Imaging compared to other analysis tasks. The diverse hardware and scanning protocols across different scanners or hospitals exacerbate this challenge. Unfortunately, the holistic model in \fl is limited to capturing general statistical patterns and lacks the specificity required for individual data sources, leading to inaccurate results. To address these problems, recent efforts focus on customized \fl methods, aiming to train local models tailored to specific data distributions. Certain methods employ hypernetworks~\cite{shamsian2021personalized} to attain personalization by creating customized weights for the target network. Nevertheless, numerous currently available hypernetwork-based Federated Learning approaches often overlook the underlying physical processes and come with significant additional training expenses.
In computer-aided diagnosis, the precision of imaging (upstream) tasks significantly impacts subsequent processes like segmentation and detection (downstream). This paper presents an in-depth \fl framework for upstream tasks, focusing on post-processing and reconstruction. Our hierarchical customized \fl framework (\hffed) for \xray Imaging addresses holistic optimization by personalizing feature adaptation and extracting invariant holistic imaging features.
In \xray Imaging, maintaining structural similarity is crucial. Our \hffed framework trains a universally shared imaging network using varied client data to capture consistent holistic characteristics. Recognizing the impact of scanning protocols and geometry parameters, \hffed introduces a hierarchical hypernetwork that dynamically adjusts imaging features using embedded knowledge.

\hffed integrates a hospital-specific hierarchical framework and a Network of Networks (\non). The hypernetwork adapts scanning parameters to customize features, enhancing invariant features of the imaging network. Utilizing \fl, it facilitates comprehensive gradient exchange for robust model training. The hospital-specific hypernetwork ensures domain-specific personalization, while \non learns universal features across diverse domains.

The contributions of this work are as follows: \\

1. Introduction of \hffed for customized \xray imaging using \fl. This is the first attempt at applying Federated Learning for customized \xray imaging.

2. The proposed \hffed framework comprises two components: the hospital-specific hierarchical hypernetwork and the Network of Networks (\non). The hypernetwork is designed for customization to address \noniid challenges, while \non captures invariant and stable features across diverse data distributions.

3. A flexible framework that can be easily extended to various \xray imaging tasks, including post-processing and reconstruction.

\section{Related Work}

\xray imaging technology has advanced significantly, enhancing diagnostic capabilities~\cite{lu2022m}. Reconstruction methods include sinogram filtration, iterative reconstruction (\texttt{IR}), and postprocessing. Sinogram filtration involves filtering raw or log-transformed data using adaptive techniques such as weighted least-squares with penalties~\cite{donoho2006compressed}. \texttt{IR} methods improve objective functions by incorporating prior knowledge like total variation and non-local means filters~\cite{chen2009bayesian}, along with regularization terms. Postprocessing methods offer convenience but lack flexibility without raw data. Deep Learning (\dl) has shown promise in low-dose computed tomography (\texttt{LDCT}) reconstruction with models like \texttt{RED-CNN} for denoising~\cite{chen2017low}, GAN-based approaches for super resolution~\cite{you2019ct}, and parameter-dependent frameworks (\texttt{PDF})~\cite{xia2021ct}. However, these centralized models raise privacy concerns and limit clinical applicability.

\subsection{Customized Federated Learning}

Federated Learning (\fl) adopts a decentralized approach that emphasizes data privacy while allowing models to learn from diverse data sources~\cite{shah2021model}. The traditional \fl method, FedAvg, introduced by McMahan et al.~\cite{mcmahan2017communication}, builds a comprehensive model by averaging local models from various parties. FedProx~\cite{li2020federated} improves upon FedAvg by enforcing proximity between local and global models. Li et al.~\cite{li2021model} introduced the model-contrastive (\texttt{MOON}) technique, which minimizes contrastive distances between local and global models. \fl's privacy-preserving features have been applied in various medical tasks~\cite{kaissis2020secure}.

In the realm of medical imaging advancements, Guo et al. \cite{guo2021multi} introduced an intermediate latent feature alignment approach for \texttt{MRI} reconstruction, requiring knowledge of target domain data, which poses practical challenges. Dinh et al. \cite{dinh2022new} addressed non-iid challenges in multitask learning within federated learning (\fl), aiming for a universally applicable model but facing significant hurdles. Zhang et al. \cite{zhang2023semi} proposed a semi-asynchronous \fl framework for short-term solar power forecasting, showing robust performance. Liang et al. \cite{liang2020think} introduced \texttt{LG-FedAvg}, linking learning representation and \fl, primarily for high-level tasks, potentially less seamless in low-level imaging assignments. \texttt{Ditto} \cite{li2021ditto}, a comprehensive-regularized multitask learning framework, integrates local and holistic models effectively, showing promising outcomes.

Ma et al. \cite{ma2022layer} introduced an approach to assess layer significance across diverse clients for tailored model aggregation. Li et al. \cite{li2021fedbn} presented \texttt{FedBN}, integrating local batch normalization (\texttt{BN}) to mitigate feature shift and achieve personalization in local hospital settings. Shamsian et al. \cite{shamsian2021personalized} proposed \texttt{pFedHN}, using a hypernetwork for local model parameter generation, particularly effective in simple models. Ashraf et al. \cite{ashraf2024transfed} explored network customization for transformer-based client-side models, limiting broad applicability. Chen et al. developed cyclic knowledge distillation for extracting semantic features from local models, focusing on classification tasks aligning semantic features across clients. Hanzely \cite{hanzely2020federated} incorporated regularization to minimize optimization gaps in federated models. Feng et al. \cite{feng2022specificity} proposed an \texttt{MRI} denoising model with shared encoders, addressing non-iid challenges in \xray reconstruction due to protocol sensitivity. Here, we introduce a hierarchical \xray reconstruction framework adaptable for diverse imaging tasks.

\section{Proposed Method}
\subsection{Problem Statement}

In the context of \xray reconstruction, the general optimization problem can be expressed as follows:

\begin{equation}
    \min_a \frac{1}{2} \lVert Xa - b \rVert_2^2 + Y(a)
\end{equation}

In this context, $\lVert\cdot\rVert_2^2$ signifies the L2 norm, and $X$ symbolizes the system matrix. The variables $a$ and $b$ denote the image for reconstruction and the measured data, respectively. The regularization term $Y(\cdot)$ integrates prior knowledge. In customized Federated Learning (\fl), hospitals strive to attain customized models from local data, improving performance through collaborative \fl methods, such as sharing specific layers. If there are $K$ hospitals, the learning process can be expressed as:
\[
\min_{\theta_1, \ldots, \theta_K} \sum_{k=1}^{K} W_k \mathbb{E}_{(a_l, b, a_n) \sim D_k} \left\| F_k(a_l, b, \theta_k) - a_n \right\|_2^2
\]

In this formulation, $F^k$ denotes the optimization model in the $k$-th hospital, $W_k$ represents the weight of the $k$-th hospital in holistic optimization, and $D_k$ signifies the dataset in the $k$-th hospital. The aim of customized Federated Learning is to discover optimal local models for different hospitals without sharing raw data.
\begin{figure}[!h]
    \centering
    \includegraphics[width = 7 cm, height = 3.5 cm]{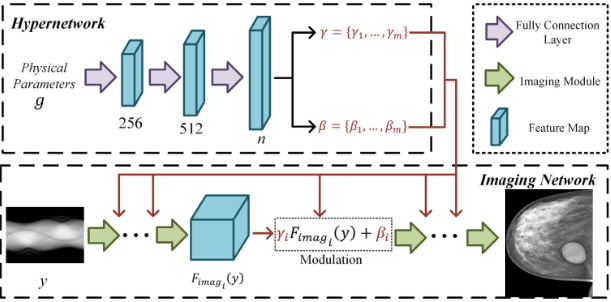}
    \caption{The proposed \hffed architecture consists of a globally shared imaging network and a hospital-specific, hierarchically-driven hypernetwork.}
    \label{img1}
\end{figure}
\subsection{\hffed Architecture}
\label{sec}

Given the substantial impact of scanning protocol and geometry parameters on the physical \xray reconstruction process, it is reasonable to leverage this knowledge for guiding the network in predicting normal \xray images. Motivated by this concept, we introduce the Hierarchical Framework-based Federated Learning (\hffed), which consists of a hospital-specific hypernetwork and a holistically shared imaging framework.
The complete assimilation process of the current imaging framework, denoted as \( L_{\text{imag}}(a, b, w) \), involves feeding inputs into the model \( L_{\text{imag}} \) with parameters \( w \), without any constraint. While assuming a uniform distribution of inputs can reduce complexity, meeting this assumption in real scenarios is challenging, leading to the non-iid problem. 
To tackle this, the hospital-specific hypernetwork functions as a monitor, adjusting the process. As a result, the assimilation process of \hffed is redefined as \( L_{\text{imag}}(a, b, w, \theta) \), where \( \theta \) denotes the result from the hypernetwork \( H(\cdot) \) constrained with \( \xi \).
In particular, the vector $n$, encompassing crucial scanning and geometry parameters, is input into $H(\cdot)$ to generate the sets of scaling factors $\gamma$ and biases $\beta$. These sets are employed to adjust the features of $F_{\text{imag}}$, and this procedure is expressed as $\gamma, \beta = H(n, \xi)$. Some elements in $n$ with large magnitudes, such as the numbers of views and detector bins, and the photon number of incident \xray, undergo normalization and logarithmization. The normalization is defined as 
\begin{equation}
    n_j = \frac{n_j - \min(n_j)}{\max(n_j) - \min(n_j)}
\end{equation}

Figure~\ref{img1} illustrates the architecture of our proposed \hffed. The structure of the hypernetwork is closely tied to the imaging network and is designed as a three-layer fully-connected network in this article. It generates scaling factors $\gamma$ and biases $\beta$ based on a 7-dimensional physical parameter vector $n$. The output sizes of the two subsequent linear layers are 256 and 512, respectively, with the final output dimensionality dependent on the imaging network's feature dimension, which is twice the imaging feature dimension.

To enhance parameter efficiency, \texttt{RED-CNN} adopts a strategy where the encoder and decoder layers share two groups of modulation factors. Similarly, in \texttt{LEARN}, all \texttt{IR} blocks share the same modulation factors. The modulation function is represented as:
\[ L^{\ast}_{\text{imag}}(b) = \gamma L_{\text{imag}}(b) + \beta. \]
This operation is applied to feature maps from various modules in the imaging network, resulting in the modification of equation (6):
\[ L^{\ast}_{\text{imag}_i}(b) = \gamma_i L_{\text{imag}_i}(b) + \beta_i, \]
where $L_{\text{imag}_i}(b)$ and $L^{\ast}_{\text{imag}_i}(b)$  represent the feature map from the $i$th module and its modulated counterpart, respectively. $\gamma_i$ and $\beta_i$ signify the regularization factor and bias for the $i$th module.

As highlighted earlier, the imaging network in \hffed is adaptable for various tasks, with imaging units varying based on different imaging methods. For instance, convolution layers serve as imaging units for postprocessing methods like \texttt{RED-CNN}~\cite{chen2009bayesian}. Conversely, for unrolled iteration methods such as \texttt{LEARN}~\cite{chen2018learn}, the imaging unit represents an unrolled iteration module.

\subsection{Implementation of NoN (Network of Networks)}

In this research, we propose \hffed, a hypernetwork-based federated learning framework designed to tackle the non-iid challenge in \xray image reconstruction. Similar to \texttt{FedAvg} and \texttt{FedProx}, \hffed involves local updates for both the hypernetwork and imaging network, with only the imaging network's updates aggregated on the server. Unlike \texttt{FedBN}, which normalizes data globally, \hffed adapts by performing local normalization due to varied \xray data distributions from different machines, challenging \texttt{FedBN's} assumptions~\cite{lim2017enhanced}. To address this, we introduce a hypernetwork to modulate feature maps within the imaging network, enabling hierarchical-driven self-normalization

Privacy and security challenges in inter-hospital collaboration are addressed by \hffed, ensuring that local data remains private. Each hospital refines its local model by minimizing the loss:
\[
\ell_k = \frac{1}{2} \mathbb{E}_{(a_i, n, b) \sim D_k} \left[ L_k(\delta_k, b, n) - a_i \right]^2
\]
where \( L_k(\cdot) \) represents the local network at the \( k \)-th hospital, parameterized by \( \delta_k \). Optimization of parameters \( w_k \) and \( \xi_k \) of \( L_{\text{imag}} \) and \( H(\cdot) \) at the \( k \)-th hospital is conducted iteratively:
\[
w^{w+1}_k, \xi^{w+1}_k = w^w_k, \xi^w_k - \lambda \nabla_{w^w_k, \xi^w_k} L^k
\]
where \( \lambda \) denotes the learning rate.
After training epochs, gradients from the imaging network are sent to the server for aggregation, while the hypernetwork remains locally for modulation.

\begin{table*}[t]
  \caption{HF-Fed's SSIM scores for the post-processing task}
  \label{tab:1}
  \centering
  \begin{adjustbox}{width=\textwidth}
    \begin{tabular}{l|S|S|S|S|S|S|S}
      \toprule
      & \multicolumn{7}{c}{SSIM Scores} \\
      \cmidrule(lr){2-8} 
      \# Models & {w/o FL} & {Fedavg} & {FedProx} & {FedBN} & {Ditto} & {pFedHN} & {HF-Fed}\\
      \midrule
      Hospital \#1 & 94.55$\pm$0.15 & 92.21$\pm$0.08 & 92.94$\pm$0.13 & 93.68$\pm$0.13 & 93.91$\pm$0.17 & 95.25$\pm$0.11 & \textbf{96.87$\pm$0.12}  \\
      Hospital \#2 & 90.42$\pm$4.22 & 94.28$\pm$4.23 & 91.85$\pm$1.50 & 91.23$\pm$1.93 & 93.02$\pm$0.88 & 90.20$\pm$0.95 & \textbf{94.64$\pm$0.22}  \\
      Hospital \#3 & \textbf{94.86$\pm$0.99} & 94.01$\pm$0.12 & 92.41$\pm$0.16 & 95.70$\pm$0.14 & 91.23$\pm$0.32 & 91.72$\pm$0.16 & 91.19$\pm$0.18  \\
      Hospital \#4 & 92.26$\pm$0.61 & 93.57$\pm$0.52 & 91.45$\pm$0.87 & 92.13$\pm$0.67 & 93.08$\pm$0.72 & \textbf{94.94$\pm$0.91} & 93.25$\pm$0.77  \\
      Hospital \#5 & 91.75$\pm$0.22 & 94.46$\pm$0.11 & 94.25$\pm$0.69 & 91.13$\pm$0.84 & 94.61$\pm$0.45 & 95.05$\pm$0.47 & \textbf{96.63$\pm$0.98}  \\
      \midrule
      Average & 92.57$\pm$1.65 & 93.11$\pm$1.65 & 92.58$\pm$1.65 & 92.70$\pm$1.65 & 93.57$\pm$1.65 & 93.03$\pm$1.65 & \textbf{95.12$\pm$1.65}  \\
      \textbf{STD} & 1.65 & 0.91 & 1.22 & 1.63 & 1.23 & 2.04 & 1.62  \\
      \bottomrule
    \end{tabular}%
  \end{adjustbox}
\end{table*}

\section{Experiments and Results}

The "\texttt{RSNA Screening Mammography Breast Cancer Detection}" dataset~\cite{rsnabreastcancerdetection} includes 54.7K \xray mammogram images from 1,000 patients, split into 4,000, 1,000, and 1,000 images for training, validation, and testing respectively. No overlap exists between these sets, with validation occurring every 200 rounds during training. The dataset is divided into five groups on average, resulting in diverse client-specific data distributions, posing a severe \texttt{non-iid} challenge. Evaluation uses the correlation coefficient (\texttt{CC}) to assess denoised image quality.
Each hospital handles a single data type with strict transmission restrictions. Post-processing uses \texttt{RED-CNN} \cite{chen2009bayesian}, and reconstruction employs \texttt{LEARN} \cite{chen2018learn} with a learning rate of $1 \times 10^{-4}$. \texttt{RED-CNN} undergoes 3 epochs over 1000 rounds, while \texttt{LEARN} uses 200 rounds. \texttt{\hffed} is compared with FedAvg \cite{mcmahan2017communication}, \texttt{FedProx} \cite{li2020federated}, \texttt{FedBN} \cite{li2021fedbn}, Ditto \cite{li2021ditto}, \texttt{pFedHN} \cite{shamsian2021personalized}, and original models without federated learning (w/o  \fl). \texttt{FedProx} applies a penalty constant of $1 \times 10^{-4}$, and \texttt{FedBN} adds personalized BN layers post-convolution. Optimization uses Adam \cite{kingma2014adam} with Mean Squared Error (\texttt{MSE}) loss. Experiments are conducted on \texttt{NVIDIA GTX} 3090 and \texttt{A100} GPUs using \texttt{PyTorch}.

\subsection{Experiments on the Large-Scale Training Set}

In this section, we compare \fl-based methods and original imaging models without \fl, using a large-scale local dataset with \texttt{RED-CNN} as the backbone network. Hyperparameters remain consistent, except each hospital has 200 images in the training set. Table~\ref{tab:1} demonstrates that \fl-based methods significantly improve performance with larger datasets, mitigating the \texttt{non-iid} problem. \hffed consistently delivers high performance across different dataset sizes, enhancing imaging quality effectively. All methods benefit from larger training samples, with \hffed remaining competitive. Our hypernetwork uses \xray geometry parameters to modulate feature maps, balancing stability and imaging performance, proving effective in achieving consistent and competitive results
\begin{figure}[t]
  \centering
  \begin{minipage}[b]{0.5\textwidth}
    \centering
    \begin{tabular}{ccc}
      \toprule
      Model & PSNR & SSIM \\
      \midrule
      W/o FL "†" & 36.58 & 0.9384 \\
      W/o FL "‡" & 38.15 & 0.9348 \\
      FedAvg "†" & 35.51 & 0.8726 \\
      FedAvg "‡" & 36.85 & 0.9158 \\
      "HF-Fed $\diamondsuit$" & 38.62 & 0.9644 \\
      "HF-Fed $\star$" & 38.95 & 0.9667 \\
      "HF-Fed $\circ$" & 39.51 & 0.9691 \\
      \midrule
      HF-Fed & \textbf{39.95} & \textbf{0.9654} \\
      \bottomrule
    \end{tabular}
    \caption*{Table 2: Results of Ablation Studies}
    \label{tab2}
  \end{minipage}%
  \begin{minipage}[b]{0.5\textwidth}
    \centering
    \includegraphics[width=\textwidth]{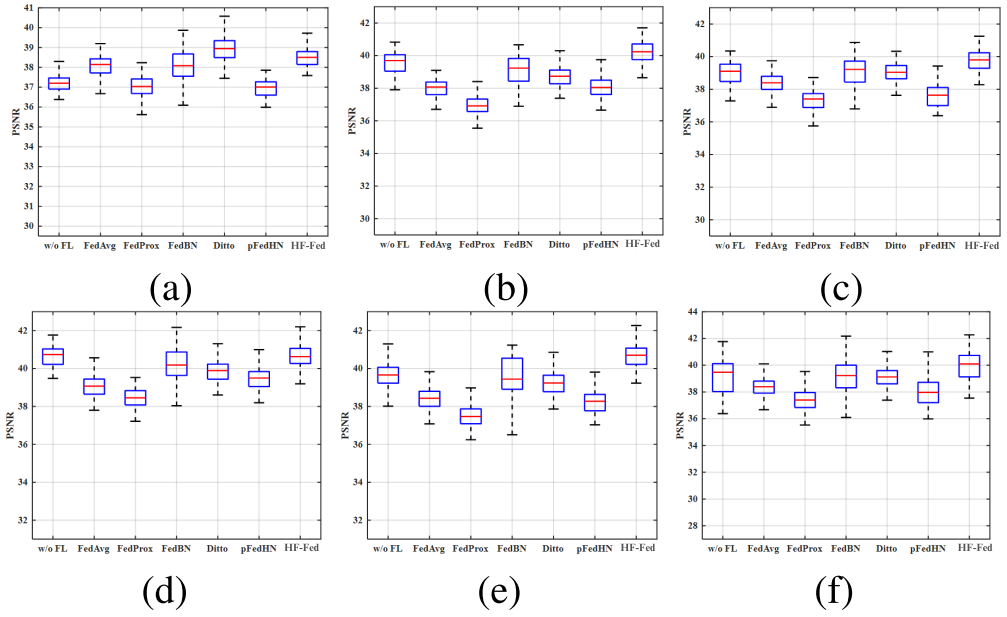}
    \caption*{Fig. 2: Boxplots of PSNR scores for post-processing}
    \label{fig2}
  \end{minipage}
\end{figure}

\subsection{Ablation Study}

In this section, we conduct ablation studies to highlight the effectiveness of various components in \hffed. The experiments follow the settings outlined in Section \ref{sec}, with results summarized in Table \textcolor{red}{2}. "†" and "‡" denote imaging networks without and with the hypernetwork, respectively, showing the significant role of the hypernetwork in improving imaging performance by addressing domain gap issues. Further, we evaluate our learning strategy by aggregating only the hypernetwork in rounds labeled "\hffed $\diamondsuit$," addressing the challenge of heterogeneous scanning parameters. Additional experiments explore the modulation scope of the hypernetwork, where "\hffed $\star$" and "\hffed $\circ$" scenarios focus on encoder and decoder modulation, respectively. Figure \textcolor{red}{2} shows the boxplots of \texttt{PSNR} based on w/o \fl, \texttt{FedAvg}, \texttt{FedProx}, \texttt{FedBN}.
\texttt{Ditto}, \texttt{pFedHN}, and \hffed for the postprocessing task. Results suggest similar performances across all modulation scenarios, indicating the effectiveness of modulating all layers for consistency and generality.

\section{Conclusion}
Current \xray imaging networks typically use centralized learning (\cl), which overlooks privacy concerns and faces challenges with non-iid data due to varied scanning protocols and equipment. Introducing \hffed, a hierarchical-based hypernetwork for \xray imaging, addresses these issues by seamlessly integrating into diverse networks (\texttt{CNN}-based, \texttt{IR}-based, transformer-based). \hffed features hospital-specific hierarchical-driven hypernetworks and a shared Network of Networks (\non), which adaptively extract stable imaging features across domains. Ablation experiments demonstrate that while \fl-absent methods achieve customized \xray reconstruction, they suffer from detail loss with limited data. In contrast, \fl-based \hffed maintains stability, recovering more details across hospitals and achieving superior reconstruction quality.

\subsubsection{Disclosure of Interests.} The authors have no competing interests to declare relevant to this article's content.

\bibliographystyle{splncs04}
\bibliography{refs}

\begin{thebibliography}{10}
\providecommand{\url}[1]{\texttt{#1}}
\providecommand{\urlprefix}{URL }
\providecommand{\doi}[1]{https://doi.org/#1}

\bibitem{ashraf2024transfed}
Ashraf, T., Bin Afzal~Mir, F., Gillani, I.A.: Transfed: A way to epitomize focal modulation using transformer-based federated learning. In: Proceedings of the IEEE/CVF Winter Conference on Applications of Computer Vision. pp. 554--563 (2024)

\bibitem{rsnabreastcancerdetection}
Chen, C.C.R.B.S.D.Y.: Rsna screening mammography breast cancer detection (2022), \url{https://kaggle.com/competitions/rsna-breast-cancer-detection}

\bibitem{chen2018learn}
Chen, H., Zhang, Y., Chen, Y., Zhang, J., Zhang, W., Sun, H., Lv, Y., Liao, P., Zhou, J., Wang, G.: Learn: Learned experts’ assessment-based reconstruction network for sparse-data ct. IEEE transactions on medical imaging  \textbf{37}(6),  1333--1347 (2018)

\bibitem{chen2017low}
Chen, H., Zhang, Y., Kalra, M.K., Lin, F., Chen, Y., Liao, P., Zhou, J., Wang, G.: Low-dose ct with a residual encoder-decoder convolutional neural network. IEEE transactions on medical imaging  \textbf{36}(12),  2524--2535 (2017)

\bibitem{chen2009bayesian}
Chen, Y., Gao, D., Nie, C., Luo, L., Chen, W., Yin, X., Lin, Y.: Bayesian statistical reconstruction for low-dose x-ray computed tomography using an adaptive-weighting nonlocal prior. Computerized Medical Imaging and Graphics  \textbf{33}(7),  495--500 (2009)

\bibitem{dinh2022new}
Dinh, C.T., Vu, T.T., Tran, N.H., Dao, M.N., Zhang, H.: A new look and convergence rate of federated multitask learning with laplacian regularization. IEEE Transactions on Neural Networks and Learning Systems  (2022)

\bibitem{donoho2006compressed}
Donoho, D.L.: Compressed sensing. IEEE Transactions on information theory  \textbf{52}(4),  1289--1306 (2006)

\bibitem{feng2022specificity}
Feng, C.M., Yan, Y., Wang, S., Xu, Y., Shao, L., Fu, H.: Specificity-preserving federated learning for mr image reconstruction. IEEE Transactions on Medical Imaging  (2022)

\bibitem{geras2019artificial}
Geras, K.J., Mann, R.M., Moy, L.: Artificial intelligence for mammography and digital breast tomosynthesis: current concepts and future perspectives. Radiology  \textbf{293}(2),  246--259 (2019)

\bibitem{guo2021multi}
Guo, P., Wang, P., Zhou, J., Jiang, S., Patel, V.M.: Multi-institutional collaborations for improving deep learning-based magnetic resonance image reconstruction using federated learning. In: Proceedings of the IEEE/CVF Conference on Computer Vision and Pattern Recognition. pp. 2423--2432 (2021)

\bibitem{hanzely2020federated}
Hanzely, F., Richt{\'a}rik, P.: Federated learning of a mixture of global and local models. arXiv preprint arXiv:2002.05516  (2020)

\bibitem{hasegawa1990physics}
Hasegawa, B.H.: The physics of medical x-ray imaging  (1990)

\bibitem{ikuta2022deep}
Ikuta, M., Zhang, J.: A deep convolutional gated recurrent unit for ct image reconstruction. IEEE Transactions on Neural Networks and Learning Systems  (2022)

\bibitem{kaissis2020secure}
Kaissis, G.A., Makowski, M.R., R{\"u}ckert, D., Braren, R.F.: Secure, privacy-preserving and federated machine learning in medical imaging. Nature Machine Intelligence  \textbf{2}(6),  305--311 (2020)

\bibitem{kingma2014adam}
Kingma, D.P., Ba, J.: Adam: A method for stochastic optimization. arXiv preprint arXiv:1412.6980  (2014)

\bibitem{li2021model}
Li, Q., He, B., Song, D.: Model-contrastive federated learning. In: Proceedings of the IEEE/CVF conference on computer vision and pattern recognition. pp. 10713--10722 (2021)

\bibitem{li2021ditto}
Li, T., Hu, S., Beirami, A., Smith, V.: Ditto: Fair and robust federated learning through personalization. In: International Conference on Machine Learning. pp. 6357--6368. PMLR (2021)

\bibitem{li2020federated}
Li, T., Sahu, A.K., Zaheer, M., Sanjabi, M., Talwalkar, A., Smith, V.: Federated optimization in heterogeneous networks. Proceedings of Machine learning and systems  \textbf{2},  429--450 (2020)

\bibitem{li2021fedbn}
Li, X., Jiang, M., Zhang, X., Kamp, M., Dou, Q.: Fedbn: Federated learning on non-iid features via local batch normalization. arXiv preprint arXiv:2102.07623  (2021)

\bibitem{liang2020think}
Liang, P.P., Liu, T., Ziyin, L., Allen, N.B., Auerbach, R.P., Brent, D., Salakhutdinov, R., Morency, L.P.: Think locally, act globally: Federated learning with local and global representations. arXiv preprint arXiv:2001.01523  (2020)

\bibitem{lim2017enhanced}
Lim, B., Son, S., Kim, H., Nah, S., Mu~Lee, K.: Enhanced deep residual networks for single image super-resolution. In: Proceedings of the IEEE conference on computer vision and pattern recognition workshops. pp. 136--144 (2017)

\bibitem{lu2022m}
Lu, Z., Xia, W., Huang, Y., Hou, M., Chen, H., Zhou, J., Shan, H., Zhang, Y.: M 3 nas: Multi-scale and multi-level memory-efficient neural architecture search for low-dose ct denoising. IEEE Transactions on Medical Imaging  \textbf{42}(3),  850--863 (2022)

\bibitem{ma2022layer}
Ma, X., Zhang, J., Guo, S., Xu, W.: Layer-wised model aggregation for personalized federated learning. In: Proceedings of the IEEE/CVF conference on computer vision and pattern recognition. pp. 10092--10101 (2022)

\bibitem{mcmahan2017communication}
McMahan, B., Moore, E., Ramage, D., Hampson, S., y~Arcas, B.A.: Communication-efficient learning of deep networks from decentralized data. In: Artificial intelligence and statistics. pp. 1273--1282. PMLR (2017)

\bibitem{narayanan2008robust}
Narayanan, A., Shmatikov, V.: Robust de-anonymization of large sparse datasets. In: 2008 IEEE Symposium on Security and Privacy (sp 2008). pp. 111--125. IEEE (2008)

\bibitem{rodriguez2019detection}
Rodr{\'\i}guez-Ruiz, A., Krupinski, E., Mordang, J.J., Schilling, K., Heywang-K{\"o}brunner, S.H., Sechopoulos, I., Mann, R.M.: Detection of breast cancer with mammography: effect of an artificial intelligence support system. Radiology  \textbf{290}(2),  305--314 (2019)

\bibitem{10.1093/jnci/djy222}
Rodriguez-Ruiz, A., Lång, K., Gubern-Merida, A., Broeders, M., Gennaro, G., Clauser, P., Helbich, T.H., Chevalier, M., Tan, T., Mertelmeier, T., Wallis, M.G., Andersson, I., Zackrisson, S., Mann, R.M., Sechopoulos, I.: {Stand-Alone Artificial Intelligence for Breast Cancer Detection in Mammography: Comparison With 101 Radiologists}. JNCI: Journal of the National Cancer Institute  \textbf{111}(9),  916--922 (03 2019). \doi{10.1093/jnci/djy222}, \url{https://doi.org/10.1093/jnci/djy222}

\bibitem{schwarz2019identification}
Schwarz, C.G., Kremers, W.K., Therneau, T.M., Sharp, R.R., Gunter, J.L., Vemuri, P., Arani, A., Spychalla, A.J., Kantarci, K., Knopman, D.S., et~al.: Identification of anonymous mri research participants with face-recognition software. New England Journal of Medicine  \textbf{381}(17),  1684--1686 (2019)

\bibitem{shah2021model}
Shah, S.M., Lau, V.K.: Model compression for communication efficient federated learning. IEEE Transactions on Neural Networks and Learning Systems  (2021)

\bibitem{shamsian2021personalized}
Shamsian, A., Navon, A., Fetaya, E., Chechik, G.: Personalized federated learning using hypernetworks. In: International Conference on Machine Learning. pp. 9489--9502. PMLR (2021)

\bibitem{slovis2002alara}
Slovis, T.L.: The alara concept in pediatric ct: myth or reality? Radiology  \textbf{223}(1), ~5--6 (2002)

\bibitem{xia2021ct}
Xia, W., Lu, Z., Huang, Y., Liu, Y., Chen, H., Zhou, J., Zhang, Y.: Ct reconstruction with pdf: Parameter-dependent framework for data from multiple geometries and dose levels. IEEE Transactions on Medical Imaging  \textbf{40}(11),  3065--3076 (2021)

\bibitem{you2019ct}
You, C., Li, G., Zhang, Y., Zhang, X., Shan, H., Li, M., Ju, S., Zhao, Z., Zhang, Z., Cong, W., et~al.: Ct super-resolution gan constrained by the identical, residual, and cycle learning ensemble (gan-circle). IEEE transactions on medical imaging  \textbf{39}(1),  188--203 (2019)

\bibitem{zhang2023semi}
Zhang, W., Chen, X., He, K., Chen, L., Xu, L., Wang, X., Yang, S.: Semi-asynchronous personalized federated learning for short-term photovoltaic power forecasting. Digital Communications and Networks  \textbf{9}(5),  1221--1229 (2023)

\end{thebibliography}

\clearpage
\setcounter{page}{1}
\renewcommand\thefigure{S\arabic{figure}}
\setcounter{figure}{0}
\renewcommand\thetable{S\arabic{table}}
\setcounter{table}{0}
\appendix

\end{document}